\documentstyle[11pt,amstex,amsfonts,amssymb,righttag]{amsart}
\numberwithin{equation}{section}
\newtheorem{Theorem}{Theorem}[section]
\newtheorem{Property}{Property}[section]

\theoremstyle{definition}

\newtheorem{Remark}{Remark}[section]
\newtheorem{Example}{Example}[section]

\title[On the martingale]{On the martingale-fair index of return
for investment funds}
\date{\today}
\thanks{Supported by the KBN. Grant  No. 1H02B 018 14.}
\author[L. Gajek and M. Ka\l uszka]{Les\l aw Gajek and Marek Ka\l uszka}
\address{Institute of Mathematics, Technical University of \L \'od\'z,
W\'olcza\'nska 215, 93-005 \L \'od\'z, Poland}
\email {kaluszka@@p.lodz.pl}
\email {gal@@p.lodz.pl}
\subjclass{\ }
\date{March 12, 2004}

\begin{document}
\maketitle
\date{March 12, 2004}
\begin{abstract}
A concept of martingale-fair index of return, consistent with Arbitrage Free
Pricing Theory, is introduced. An explicit formula for the
 average rate of return of a group of investment/pension funds in a discrete time
stochastic model is derived and several properties of this index are shown.
In particular, it is proven to be martingale-fair,
i.e. be  a martingale
provided the prices of assets on the financial market form a vector
martingale.
The problem of merger of the funds is treated
in detail.
\end{abstract}

\section{Introduction}

Statistical indexes play a very important role in both economic theory and
practice for they simply aggregate information coming from the economy to every decision maker.
There are two modern approaches to selecting an index number formula. The axiomatic approach
--- dating back to Fischer (1922) --- focuses on a desired perfomance of the index in response
to particular types of changes. The alternative approach, including e.g. cost of living index theory,
concerns index's ability to reflect a substitution behavior on the part of economic agents.

This paper concerns both approaches applied to selecting an index  formula
which would reflect accurately investment results of investment pension funds.
We investigate an index formula
which accounts for the rate of return achieved by the group of funds as well as satisfies
several axioms the most important of which is the martigale property. This axiom is based
on the Arbitrage Free Pricing Theory due to S. Ross (see \cite{Ro}).
According to his Fundamental Theorem of Asset Pricing the securities market is arbitrage free
if and only if there exists a risk neutral probability measure (see Panjer et al (1998) p. 211).
But then the discounted value of the portfolio corresponding to a self-financing strategy is a martigale
under a risk-neutral probability measure (see e.g. Proposition 5.3.8 of Panjer et al (1998)).
This shows a central role of martingales in modelling securites market
 as well as the importance of the notion of martingale itself.

An important consequence of the Fundamental Theorem of Asset Pricing is that any index
 properly reflecting investment efficiency of the group
 of investment funds should behave like a martingale if the whole market value is a martingale.
Throughout the paper we call this property the {\em martingale fairness axiom}
 (shortly, the fairness
axiom) and say shortly that the index is {\em fair}. To enlight the name of the axiom,
recall that a random game with two players is called {fair} if the expectation of its
 pay off is zero.
Martingale fairness  means that a multiperiod game has, after each period, a pay off with
a conditional expectation zero subject to a given result from the previous period.
Thus it may be specially useful in testing stochastic dynamics of financial
market indexes.

The problem of measuring investment results is also of great practical importance.
In several countries statistical indexes of such kind are used as benchmarks for
privately managed pension funds.
Using ``naive'' indexes may lead then to overestimation of the market investment performance
increasing the risk of an extra contribution to the fund's assets by this assets
management company which operates the fund performing below the benchmark (see e.g. \cite{GaOs}).
Thus the benchmark should be fair in the sense that it reflects properly the
market performance. The martingale --- fairness axiom, formulated in this paper,
formalizes that intuition.
Some other arguments to apply carefully chosen investment benchmarks for pension funds
can be found in \cite{Am}.

To be more concrete, let us consider a group of $n$ investment funds.
Let $k_{i}(t)$ and
$w_{i}(t)$ denote a number of all accounting units possessed by the members of the
$i$-th investment fund and a value of the $i$-th fund unit at  time $t$,
respectively.
The problem considered in this paper is how to define an average rate of
return to meet the fairness axiom as well as several other requirements
in a discrete time stochastic model.
In Section 2 we develop a simple stochastic model with discrete time.  This
allows us to derive a proper definition of the average rate of return. Our
definition is as follows
\begin{equation}
\bar{r}_{A}(s,t) =\prod^{t-1}_{u=s}
\left(1+{\sum^{n}_{i=1} r_{i}(u,u+1)k_i(u)w_i(u)\over
\sum_{i=1}^nk_i(u)w_i(u)} \right) -1.\label{(1.1)}
\end{equation}
This model is also a
starting point to a more involved continuous time stochastic model of funds
dynamics presented in Gajek and Ka\l uszka(2004).

The paper sheds a new light on the problem of constructing
index numbers.
The axiomatic approach, exemplified by Irving Fisher's
{\it The Making of Index Numbers} (1922), evaluates indexes
based on whether they perform as desired when prices or quantities undergo particular
types of changes.
Eichorn and Voeller (1976) provided a discussion of the ``test'' approach
to index number theory. Bulk (1995) gave a survey of axiomatic price index theory
(a comprehensive account of the price index was given by Afriat (1977)).
Fisher and Shell (1972, 1992) surveyed developments in statistical
and economic index numbers
and in productivity measurement; see also Eichorn el al. (1978), Diewert (1976, 1978),
Balk and Diewert (2001) and Dumagan (2002).
The stochastic approach to the construction of price index numbers was presented
by Ogwang (1995) (see also the papers cited therein).
Comparing with the above authors, we introduce {\em martingale fairness}
--- a new axiom to test the desired performance of the rate of return index
in response to martingale-type changes of the asset prices.
The axiom concerns stochastic approach to the market dynamics, is mathematically consistent
with Arbitrage Free Pricing Theory and has appropriate economic implications.

\section{Average rate of return in a discrete-time model}
Throughout this Section we assume that $k_{i}(t)$ and $w_{i}(t)$ are
observed in discrete moments $t=0,1,2,\ldots $. The problem is
to find a proper definition of the average return for a group of $n$
investment funds during a given time period $[s,t]$. Let us denote it by
$\bar{r}_{A}(s,t)$. Clearly, $\bar{r}_{A}(s,t)$ should take values in
$(-1,\infty )$.

\subsection{The model}
We will use the following state-variables:

$c_{j}(t) =$ price of asset $j$ of the financial market at time $t$,
$j=1,\ldots, N,$

$u_{ij}(t) =$ number of units of asset $j$ possessed by the $i$-th fund at
time $t$, $i=1,\ldots  ,n$, $j=1,\ldots , N,$

$w_{i}(t) =$ value of participation unit of the $i$-th fund at
time $t$,

$k_{i}(t) =$  number of units of the $i$-th fund at time $t$,

$d_{i}(t) =$ the amount of contributions of the $i$-th fund's liabilities
minus the drawdown amount of $i$-th fund's liabilities at time $t$,

$d(t) =\sum^{n}_{i=1}d_{i}(t)$,

$A_{i}(t) =$ value of $i$-th fund's assets,

$A(t) = \sum^{n}_{i=1}A_{i}(t)$,

$A^{*}_{i}(t) =A_{i}(t)/A(t)$  --- a relative value of the
assets of the $i$-th fund at time $t$,

$\Delta f(x)=f(x+1)-f(x+)$.

\medskip
We will assume that:

1) All investments are infinitely divisible.

2) There are no transaction costs or taxes and the assets pay no
dividends.

3) Member does not pay for allocation of his/her wealth.

4) No consumption of funds exists.

\medskip
Let us consider a probability space $(\Omega ,{\cal F}, {\Bbb P})$.
Let ${\Bbb F}=\{{\cal F}_{0}$, ${\cal F}_{1}$, ${\cal F}_{2},\ldots  \}$ be a
filtration, i.e. a stream of $\sigma $-algebras of subsets of $\Omega $ with
the property ${\cal F}_{0}\subseteq {\cal F}_{1}\subseteq
{\cal F}_{2}\subseteq \ldots  \subseteq {\cal F}$.
Without loss of generality, we may put
${\cal F}_{0}=\{\emptyset ,\Omega \}$.
The
filtration ${\Bbb F}$ describes the process how information is revealed to the
investors. We will assume that $c_{i}(t)$ is measurable
with respect to ${\cal F}_{t}$ (written ${\cal F}_{t}$-measurable)
 for each $i$, $t$. Given $t$, we have
\begin{equation}
\hspace{-1cm}w_{i}(t)k_{i}(t)=u_{i1}(t)c_{1}(t)+
\ldots  +u_{iN}(t)c_{N}(t),\quad i=1,\ldots  n.\label{(2.1)}
\end{equation}
Here and subsequently, the symbol $X=Y$ means that the random
variables $X$, $Y$ are defined on $(\Omega ,{\cal F},{\Bbb P})$ and
${\Bbb P}(X=Y)=1$. We assume that each random variable $w_{i}(t)$ is
adapted to ${\Bbb F}$
means that $w_{i}(t)$ is ${\cal F}_{t}$-measurable
for each $i$, $t$. We also
assume that both $k_{i}(t)$ and $u_{ij}(t)$ are adapted
to ${\Bbb F}$.
It means that  we are allowing the investor to buy (sell) units after the
values $c_{i}(t)$ are observed.

At any  time $t$, split of units is allowed. The new price of one unit
and the number of units of $i$-th fund are denoted by $w_{i}(t+)$ and
$k_{i}(t+)$, respectively. At time $t$ we have
\begin{equation}
w_{i}(t+)k_{i}(t+)=w_{i}(t)k_{i}(t),\quad i=1,\ldots , n,\label{(2.2)}
\end{equation}
and at the time $t+1$,
\begin{multline}
w_{i}(t+1)k_{i}(t+)=u_{i1}(t)c_{1}(t+1)+
\ldots  +u_{iN}(t)c_{N}(t+1),\\
 i=1,\ldots , n.\label{(2.3)}
\end{multline}
From \eqref{(2.1)}--\eqref{(2.3)} we get
\begin{equation}
k_{i}(t+)(w_{i}(t+1)-w_{i}(t+))=
\sum^{N}_{j=1}u_{ij}(t)(c_{j}(t+1)-c_{j}(t))\label{(2.4)}
\end{equation}
for $i=1,\ldots , n$. Moreover, any member of the $i$-th fund reallocate his/her wealth.
The members withdraw $k^{(W)}_{i}(t+1)w_{i}(t+1)$
of monetary units from the $i$-th fund and invest
$k^{(I)}_{j}(t+1)w_{j}(t+1)$ of the amount
 $\sum _{i=1}^nk^{(W)}_{i}(t+1)w_{i}(t+1)$ in the $j$-th fund,
where $k^{(W)}_{i}(t+1)$ and $k^{(I)}_{i}(t+1)$ are
${\cal F}_{t+1}$-measurable random variables
for each $i$. At time $t+1$, the stream of liability payments also
changes balance of the $i$-th fund. As a consequence the number of units of
the $i$-th fund changes from  $k_{i}(t+)$ to $k_{i}(t+1)$:
\begin{align}
&k_{i}(t+1)w_{i}(t+1)=k_{i}(t+)w_{i}(t+1)-
k^{(W)}_{i}(t+1)w_{i}(t+1)
+\notag \\
\label{(2.5)}
& +k^{(I)}_{j}(t+1)w_{i}(t+1)+d_{i}(t+1),
\quad i=1,2,\ldots , n,
\end{align}
where
$d_{i}(t+1)$ is ${\cal F}_{t+1}$-measurable each $i$, $t$. After summing equations
in \eqref{(2.5)}, we get the following simple equation
\begin{equation}
\sum^{n}_{i=1}w_{i}(t+1)(k_{i}(t+1)-k_{i}(t+))=d(t+1).\label{(2.6)}
\end{equation}
After
allocation of clients, the management of the $i$-th fund rebalances the
portfolio:
\begin{align}
w_{i}(t+1)k_{i}(t+1)=&u_{i1}(t+1)c_{1}(t+1)+ \ldots +\notag \\
&+u_{iN}(t+1)c_{N}(t+1)\label{(2.7)}
\end{align}
for $i=1,\ldots ,n$.

To conclude, our mathematical model of dynamics of a group of funds leads to
the following stochastic difference equations:
\begin{align}
w_{i}(t)k_{i}(t)=&\sum^{N}_{j=1}u_{ij}(t)c_{j}(t),\label{(2.8)}\\
w_{i}(t+)k_{i}(t+)=&w_{i}(t)k_{i}(t), \label{(2.9)}\\
k_{i}(t+)\Delta w_{i}(t)=&\sum^{N}_{j=1}u_{ij}(t)
\Delta c_{j}(t),\label{(2.10)}\\
w_{i}(t+1)\Delta k_{i}(t)=&\left(k^{(I)}_{i}(t+1)-k^{(W)}_{i}(t+1)\right)w_{i}(t+1)+\notag \\
&+d_{i}(t+1),\label{(2.11)}\\
w_{i}(t+1)\Delta k_{i}(t)=&\sum^{N}_{j=1}c_{j}(t+1)
\Delta u_{ij}(t),\label{(2.12)}
\end{align}
where $t=0,1,2\ldots  $ and $i=1,2,\ldots , n$. In this model functions
$u_{ij}$, $k^{(I)}_i$ and $k^{(W)}_{i}$ play a role of control variables.

\subsection{Definition of average return}
Our definition of the average
return in discrete-time model in the time period $[s,t]$ is
given by the formula
\begin{equation}
\bar{r}_{A}(s,t) =\prod^{t-1}_{u=s}
\bigl(1+\sum^{n}_{i=1} A^{*}_{i}(u)r_{i}(u,u+1)\bigr) -1,\label{(2.13)}
\end{equation}
where $r_{i}(u,u+1)$ is the return of the $i$-th fund in the time interval
$(u,u+1]$, i.e.
\begin{equation}
r_{i}(u,u+1)={w_{i}(u+1)-w_{i}(u+)\over w_{i}(u+)}.\label{(2.14)}
\end{equation}
For convenience, we put $\bar{r}_{A}(s,s)=0$ for each $s$.

We give two main arguments for using  definition \eqref{(2.13)}. The first one
is based on analysis of changes of total assets of the funds. Observe
that
$$
{A(t)-A(s)\over A(s)}=\prod^{t-1}_{u=s}\left(1+{A(u+1)-A(u)\over A(u)}\right)-1.
$$
Applying \eqref{(2.2)}  and \eqref{(2.6)} we get
\begin{align*}
&A(u+1)-A(u)=\sum^{n}_{i=1}w_{i}(u+1)k_{i}(u+1)-
\sum^{n}_{i=1}w_{i}(u)k_{i}(u)=\\
&=\sum^{n}_{i=1}w_{i}(u+1)k_{i}(u+1)-\sum^{n}_{i=1}w_{i}(u+)k_{i}(u+)=\\
&=\sum^{n}_{i=1}w_{i}(u+1)(k_{i}(u+1)-k_{i}(u+))+
\sum^{n}_{i=1}(w_{i}(u+1)-w_{i}(u+))k_{i}(u+)=\\
&=d(u+1)+\sum^{n}_{i=1}A_{i}(u){w_{i}(u+1)-w_{i}(u+)\over w_{i}(u+)}.
\end{align*}
Hence
$${A(t)-A(s)\over A(s)}=\prod^{t-1}_{u=s}\left(1+{d(u+1)\over A(u)}
+\sum^{n}_{i=1}A^{*}_{i}(u)r_{i}(u,u+1)\right) - 1.
$$
After removing the influence of the amount of contribution on fund's
liabilities, we arrive at
$${A(t)-A(s)\over A(s)} =\bar{r}_{A}(s,t).$$

The second argument is as follows. Clearly
$$
\bar{r}_{A}(t,t+1)=\sum^{n}_{j=1}A^{*}_{j}(t) r_{j}(t,t+1).
$$
Since $A^{*}_{j}(t)\ge 0$ and $\sum^{n}_{j=1}A^{*}_{j}(t)=1$,
we have the following interpretation
of the average return: $\bar{r}_{A}(t,t+1)$ is
equal to the expected return of
$K_{0}$ monetary units chosen at random from the assets of the group of funds at
time $t$, i.e.
$$
\bar{r}_{A}(t,t+1)={\Bbb E}r_{J}(t,t+1),
$$
where $J$ is a random variable with the distribution
${\Bbb P}(J=j)=A^{*}_{j}(t)$,
$j=1,\ldots  , n$. Moreover, for every $s<t$,
\begin{align*}
\bar{r}_{A}(s,t)=&\prod^{t-1}_{u=s }\bigl((1+{\Bbb E}r_{J_{u}}(u,u+1)\bigr)-
1=\\
=&{\Bbb E}\prod^{t-1}_{u=s }\bigl((1+r_{J_{u}}(u,u+1)\bigr)- 1,
\end{align*}
where
$J_{s},\ldots  , J_{t-1}$ are independent random variables such that ${\Bbb
P}(J_{u}=j)=A^{*}_{j}(u)$, $j=1,\ldots  ,n$, $u=0,1,2,\ldots  $. This means
that if we repeat the procedure of choosing independently and sequentially
$K_{0}$ monetary units from the assets of one fund at time $s,s+1,\ldots , t-1$ (with
reinvestment at funds), then our capital at time t will be a random
variable $K$ with $$ {\Bbb E}K=K_{0}(1+\bar{r}_{A}(s,t)). $$ In other
words, the expected value of rate of return is equal to the average rate of
return:
$$
{\Bbb E}\left({K-K{ } _{0}\over K{ } _{0}}\right)=\bar{r}_{A}(s,t).
$$

\begin{Remark}\label{rem2.1}
After writting the paper we found out that
the definition
\eqref{(2.13)} appears in a quite different context in Barber et al (2001),
 but we have not found so far any analysis
justifying the use of it as a definition of the average rate of return.
\end{Remark}

\section{Martingale fairness and other properties of the average rate of return
$\bar{r}_{A}$}

The main result of the paper  is the following theorem.

\begin{Theorem} {
The index number $\bar r _A$ is fair, i.e.
if $\{c_{i}(t)$, $t=0,1,2,\ldots  \}$
is an ${\Bbb F}$-martingale for each
$i$, then  $\{\bar{r}_{A}(0,t)$,
$t=0,1,2,\ldots  \}$ is an ${\Bbb F}$-martingale. Moreover, if
$\{c_{i}(t)$, $t=0,1,2,\ldots  \}$ is an ${\Bbb F}$-submartingale (resp.
${\Bbb F}$-supermartingale) for each $i$,
then $\{\bar{r}_{A}(0,t)$, $t=0,1,2,\ldots  \}$ is an
${\Bbb F}$-submartingale (resp. ${\Bbb F}$-supermartingale)}.
\end{Theorem}

\begin{pf}
By definition,
the random variable $\bar{r}_{A}(0,t)$ is ${\cal F}_{t}$-measurable
for each $t$. By assumption both  $k_{i}(t)$ and
$w_{i}(t)$ are ${\cal F}_{t}$-measurable.
Moreover, $A^{*}_{i}(t)$ is ${\cal F}_{t}$-measurable, since
$$
A^{*}_{i}(t)={w_{i}(t)k_{i}(t)\over \sum^{n}_{i=1}w_{i}(t)k_{i}(t)}.
$$
Of course $\sum^{n}_{j=1}A^{*}_{j}(t)=1$ and we have
\begin{align*}
&{\Bbb E}\left(\bar{r}_{A}(0,t+1)\mid
{\cal F}_{t}\right)=(\bar{r}_{A}(0,t)+1){\Bbb E}
\left[\sum^{n}_{i=1}A^{*}_{i}(t) {w_{i}(t+1)\over w_{i}(t+)}
\mid {\cal F}_{t}\right]-1=\\
&=(\bar{r}_{A}(0,t)+1)\left(\sum^{n}_{i=1}A^{*}_{i}(t){\Bbb E}
\left[{w_{i}(t+1)\over w_{i}(t+)}\mid {\cal F}_{t}\right]\right)-1.
\end{align*}
We show that
$$
{\Bbb E}\left[{w_{i}(t+1)\over w_{i}(t+)}\mid {\cal F}_{t}\right]=1
$$
for each $i$. Recall that
$$w_{i}(t+1)k_{i}(t+)=u_{i1}(t)c_{1}(t+1)+
\ldots  +u_{iN}(t)c_{N}(t+1),
$$
$$
w_{i}(t+)k_{i}(t+)=w_{i}(t)k_{i}(t)=
u_{i1}(t)c_{1}(t)+\ldots  +u_{iN}(t)c_{N}(t),
$$
for $i=1,\ldots , n$.
Since $c(t)$ is an ${\Bbb F}$-martingale, we get
\begin{align*}
&{\Bbb E}\left({w_{i}(t+1)\over w_{i}(t+)}\mid {\cal F}_{t}\right)=
{1\over k_{i}(t+)w_{i}(t+)}{\Bbb E}
\left(\sum^{N}_{j=1}u_{ij}(t)c_{j}(t+1)\mid {\cal F}_{t}\right)=\\
&={1\over k_{i}(t+)w_{i}(t+)}
\sum^{N}_{j=1}u_{ij}(t){\Bbb E}(c_{j}(t+1)\mid {\cal F}_{t})=    \\
&={1\over k_{i}(t+)w_{i}(t+)}\sum^{N}_{j=1}
u_{ij}(t)c_{j}(t)={k_{i}(t)w_{i}(t)\over k_{i}(t+)w_{i}(t+)}=1.
\end{align*}
The proof of the first part of the theorem is completed. The proof
of the second part is analogous so it is omitted.
\end{pf}

\begin{Remark}\label{rem3.1}
A natural question arises if other indexes of the average rate of return are
martingale-fair.
An interesting example is provided by
 the Polish law regulations on operation of pension funds (see The Law on Organisation and
Operation of Pension Funds, Art. 173, Dziennik Ustaw Nr 139 poz. 934, Art.
173; for the English translation see {\it Polish Pension ...}, 1997). The
following definition of the average return of a group of pension funds can
be found there:
\begin{align}
\bar{r}_{PL}(s,t)=&\sum^{n}_{i=1}{1\over 2}r_{i}(s,t)
\left({w_{i}(s)k_{i}(s) \over \sum^{n}_{j=1}w_{j}(s)k_{j}(s)}+\right. \notag \\
&\left. +{w_{i}(t)k_{i}(t) \over \sum^{n}_{j=1}w_{j}(t)k_{j}(t)}\right),
\label{(2.16)}
\end{align}
where  $r_{i}(s,t)$  denotes the rate of return of the
$i$-th fund, that is, $$
r_{i}(s,t)={w_{i}(t)-w_{i}(s)\over w_{i}(s)}.$$
The average rate of return defined above
is not  martingale-fair in general. In
fact, assume that $k_{i}(s)=k$, $w_{i}(0)=1$, and $u_{ij}(s)=u_{ij}$
with $k,u_{ij}\in {\Bbb R}$ for each $i,j,s$. After an elementary algebra
we get
$$ \bar{r}_{PL}(0,t)={1\over 2n}\sum^{n}_{i=1} w_{i}(t)-1+{1\over 2}
{\sum^{n}_{i=1}(w_{i}(t)){ } ^{2} \over \sum^{n}_{i=1}w_{i}(t)}, $$
where
$w_{i}(t)=(u_{i1}c_{1}(t)+\ldots  +u_{iN}c_{N}(t))/k$. Since $(w_{1}+\ldots
+w_{n})^{2}\le n(w_{1}^2+\ldots +w_{n}^2)$ for all reals $w_{i}$, we
have
\begin{align}
{\Bbb E}\bar{r}_{PL}(0,t)=&{1\over 2n} \sum^{n}_{i=1}{\Bbb
E}w_{i}(t)-1+{1\over 2} {\Bbb E}\left({\sum^{n}_{i=1}(w_{i}(t)){ }
^{2}\over \sum^{n}_{i=1}w_{i}(t)}\right)\ge \notag \\
\ge& -{1\over 2} +{1\over
2n} {\Bbb E} \sum^{n}_{i=1}w_{i}(t)=0={\Bbb
E}\bar{r}_{PL}(0,0),\label{(2.17)}
\end{align}
and equality holds in \eqref{(2.17)} if and
only if ${\Bbb P}(w_{1}(t)=\ldots  =w_{n}(t))=1$.

Suppose that $u_{ik}\neq u_{jk}$ for some $i,j,k$,
and suppose $c_{1}(t), \ldots  , c_{N}(t)$
are not linearly dependent, i.e. for any reals $a_{1}$, $\ldots$,
$a_{N}$ such that
$\mid a_{1}\mid +\ldots  +$ $\mid a_{N}\mid \neq 0$,
$$
{\Bbb P}(a_{1}c_{1}(t)+\ldots  +a_{N}c_{N}(t)= 0)<1.
$$
Then ${\Bbb P}(w_{1}(t)=\ldots  =w_{n}(t))<1$ and
$$
{\Bbb E}\bar{r}_{PL}(0,t)>{\Bbb E}\bar{r}_{PL}(0,0).
$$
This means that $\{\bar{r}_{PL}(0,t)$,
$t=0,1,2,\ldots  \}$ cannot be a martingale.
\end{Remark}

Now, we formulate several axioms which any properly defined
average rate of return should satisfy. Some of them can be found in
Kellison(1991). The average return
$\bar{r}_{A}$ meets all the demands. The proofs
are straightforward so they will be omitted.

\begin{Property}\label{1}
 { If the group consists only of the $i$-th fund, then
$$
\bar{r}_{A}(s,t)={w_{i}(t)-w_{i}(s+) \over w_{i}(s+)}
$$}
\end{Property}

\begin{Property}[Multiplication axiom]\label{2}
 {For every $s<u<t$
$$
1+\bar{r}_{A}(s,t)=(1+\bar{r}_{A}(s,u))(1+\bar{r}_{A}(u,t))
$$}
\end{Property}

The next property says that the average rate of return $\bar{r}_{A}$ is
consistent with respect  to  grouping of funds.

\begin{Property}[Consistency in aggregation-axiom]\label{3}
{If funds are grouped, and if the average rate of return of
every group is calculated over the time interval $[t,t+1)$, then the
average rate of return of groups equals to the average rate of return of all
funds over the time interval $[s,s+1)$.}
\end{Property}

It is easy to check that $\bar{r}_{PL}$ does not possess Properties \ref{2} and \ref{3}.
For instance, we show that Property \ref{3} is not satisfied. Suppose
there are three funds with the rate of returns $r_{1}, r_{2}$ and $r_{3}$,
respectively, on a given time period $[s,s+1]$. Then
$$\bar{r}_{PL}(s,s+1)=\sum^{3}_{i=1}
{1\over 2}r_{i}\left({A_{i}(s)\over A(s)} + {A_{i}(s+1)\over A(s+1)}\right),
$$
where
$$
A(s+1)=\sum^{3}_{i=1}A_{i}(s+1).
$$
Let us group the first fund and the second one. In this way there
are two funds: the first one has the return
$$
r^{g}_{1}=\sum^{2}_{i=1}{1\over 2}r_{i}
\left({A_{i}(s)\over \sum^{2}_{i=1}A_{i}(s)} +
{A_{i}(s+1)\over \sum^{2}_{i=1}A_{i}(s+1)}\right),
$$
and the return of the second one
equals $r^{g}_{2}=r_{3}$. By  definition
\eqref{(2.16)},
$$\bar{r}^{g}_{PL}={1\over 2}r^{g}_{i}
\left({\sum^{2}_{i=1}A_{i}(s)\over A(s)} + {\sum^{2}_{i=1}A_{i}(s+1)\over
A(s+1)}\right)+ {1\over 2}r^{g}_{2}\left({A_{3}(s)\over A(s)} +
{A_{3}(s+1)\over A(s+1)}\right). $$
After an easy algebra we get
\begin{align*}
\bar{r}^{g}_{PL}=& \sum^{2}_{i=1} {1\over 4}r_{i}\left(A_{i}(s)+
A_{i}(s+1)\right)\left({1\over A(s)} + {1\over A(s+1)}\right) +\\
&+{1\over
2}r_{3}\left({A_{3}(s)\over A(s)} + {A_{3}(s+1)\over A(s+1)}\right).
\end{align*}
Clearly
$\bar{r}^{g}_{PL}\neq \bar{r}_{PL}$. For example, if we assume that the
number of units of each fund is constant and if $A_{1}(s)=10^{6}$,
$A_{1}(s+1)=1.1\cdot 10^{6}$, $A_{2}(s)=3\cdot 10^{6}$,
$A_{2}(s+1)=3.21\cdot 10^{6}$, $A_{3}(s)=4\cdot 10^{6}$,
$A_{3}(s+1)=4.19\cdot 10^{6}$, then
$$ \bar{r}_{PL}=6.26\%,\quad
\bar{r}^{g}_{PL}=7.48\%. $$
Changing $A_{3}(s+1)=4.19\cdot 10^{6}$  for
$A_{3}(s+1)=4.39\cdot 10^{6}$, we get
$$ \bar{r}_{PL}=8.76\%,\quad
\bar{r}^{g}_{PL}=7.47\%. $$
Hence  the average return
$\bar{r}_{PL}$ may increase or decrease after grouping the funds. In our opinion, the fact that
$\bar{r}_{PL}$ has not Property \ref{3} makes it useless.
Indeed, using it in Poland to report on pension funds investment efficiency leads to
several practical problems described in \cite{GaOs}.
A more detailed consideration
on grouping funds is placed in Section \ref{sec4}.

\begin{Property}\label{4}
{If on a subset of a probability space the accounting units
of all funds have the same values over $[s,t]$, i.e.
$w_{1}(u)=w_{2}(u)=\ldots =w_{n}(u)$ for every $s\le u\le t$, then $$
\bar{r}_{A}(s,t)={w_{1}(t)-w_{1}(s+)\over w_{1}(s+)}$$
holds on the same subset.}
\end{Property}

\begin{Property} \label{5}
 {Suppose that $k_i(u)=\alpha _i \phi (u)$ for
all $u\in [s,t]$, $i=1,2,\ldots, n$, where $\alpha _i\ge 0$, $\sum
_{i=1}^n\alpha _i=1$ and $\phi
: [s,t]\rightarrow (0,\infty)$. Then $$ \bar{r}_{A}(s,t)= \sum _{i=1}^n A^*_i(s)r_i(s,t), $$ where
$r_i(s,t)=(w_i(t)-w_i(s+))/w_i(s+)$.}
\end{Property}

Observe that the above formula is
equivalent to formula (12) of Gajek and Ka\l uszka (2001). If the number of
units of every fund is constant over the time interval $[s,t]$, i.e. $\phi
(u)=1$ for all $u$, then
$$ \bar{r}_{A}(s,t)= {A(t)-A(s)\over
A(s)}. $$

\begin{Property}\label{6}
{For every $s<t$
$$
\min_{1\le i\le n,s\le u<t}r_{i}(u,u+1)\le
\bar{r}_{A}(s,t)\le \max_{1\le i\le n,s\le u<t}r_{i}(u,u+1).
$$}
\end{Property}

The next property says that the influence of small funds on the
average return of the group of funds is asymptotically negligible.

\begin{Property}\label{7}
{Given  $k$ and a fixed elementary event, if $\max_{i\neq k}A_{i}(u)\le
\theta A_{k}(u)$, $u\in [s,t)$, then
$$
 \bar{r}_{A}(s,t)= {w_{k}(t)-w_{k}(s)\over w_{k}(s)}+\delta (\theta), $$
 where $\delta(\theta)\to 0$ as $\theta\to 0$.}
\end{Property}

Observe that the following definition of average rate of return does
not possess Property \ref{7}:
$$
\bar{r}_V(s,t)=[(1+r_{1}(s,t))\cdot \ldots \cdot (1+r_{n}(s,t))]^{1/n}-1,
$$
where $r_{i}(s,t)$ is defined by \eqref{(2.14)}.
The average return $\bar{r}_V(s,t)$ is a
counterpart of the well-known Value Line Composite Index (VLIC
index) since
$$
\bar{r}_V(s,t)=\left(\prod _{i=1}^n { w_i(t)\over w_i(s)}\right)^{1/n}-1.
$$
The next property says that if we move some assets from a less effective
fund to a more effective one, then the average rate of return increases.

\begin{Property}\label{8}
{Let $s<u<$t.
Suppose that $w_{i}(s)=w_{i}(u)=w_{i}(t)$ for every
$i=3,4,\ldots  , n$ and suppose $r_{1}(s,u)<r_{2}(s,u)$.
Moreover, suppose that
some clients transfer their assets from the first fund to the second
one at time $u$. Then the average return increases over the time
interval $[s,t]$ if and only if $r_{1}(u,t)<r_{2}(u,t)$.}
\end{Property}

\section{Merger of funds}\label{sec4}

Suppose that there
exists $n$ pension funds at time  $t=0,1,2,\ldots  , \tau $.
At time $\tau  $ the $n$-th fund
and the $(n-1)$-th one form a new fund, say $(n-1)$-th. The assets of the
new fund are equal to
$$ k_{n-1}(\tau )w_{n-1}(\tau )+k_{n}(\tau )w_{n}(\tau).
$$
At time $\tau $, the  number of
units of the new fund will be denoted by $k_{n-1}(\tau +)$. The value of one unit of the
new fund, $w_{n-1}(\tau +)$, is calculated according to the formula
$$ k_{n-1}(\tau +)w_{n-1}(\tau +)=k_{n-1}(\tau )w_{n-1}(\tau )+
k_{n}(\tau )w_{n}(\tau ). $$
Suppose that up to  time $T>\tau $, the
number of funds is constant. How to calculate the average return over the time period
$[0,T]$?

Observe that the merger of the above two funds can be treated as an
allocation of all assets of the $n$-th fund to the $(n-1)$-th one. After
allocation, the units of the $(n-1)$-th fund split so that the
new value of the unit of the $(n-1)$-th fund is equal to
$w_{n-1}(\tau +)$ and the number of units is equal to
$k_{n-1}(\tau +)$. By
\eqref{(2.13)},
\begin{align}
 \bar{r}_{A}(0,T)=& \prod^{\tau -1}_{t=0
}\left(\sum^{n}_{j=1}A^{*}_{j}(t) {w_{j}(t+1)\over w_{j}(t)}\right)\times \label{(2.18)} \\
&\times \left(\sum^{n-2}_{j=1}A^{*}_{j}(\tau ){w_{j}(\tau +1)\over
w_{j}(\tau )}+ A^{*}_{n-1}(\tau ){w_{n-1}(\tau +1)\over w_{n-1}(\tau
+)}\right)\times \notag \\
&\times \prod^{T-1}_{t=\tau
+1}\left(\sum^{n-2}_{j=1}A^{*}_{j}(t) {w_{j}(t+1)\over
w_{j}(t)}+A^{*}_{n-1}(t){w_{n-1}(t+1) \over w_{n-1}(t)}\right)-
1,\notag
\end{align}
provided there is no split of units of others funds up
to the time $T$. The rate of return $r'_{n-1}(0,\tau )$ of the new
$(n-1)$-th fund at the moment $\tau $ equals
\begin{align}
r'_{n-1}(0,\tau )
=&\prod^{\tau -1}_{t=0}\left({A_{n-1}(t) \over A_{n-1,n}(t)}
{w_{n-1}(t+1)\over w_{n-1}(t)} +\right. \notag \\
&\left. +{A_{n}(t)\over A_{n-1,n}(t)}
{w_{n}(t+1)\over w_{n}(t)}\right) - 1, \label{(2.19)}
\end{align}
where by
$$
A_{n,n-1}(t)=k_{n-1}(t)w_{n-1}(t)+k_{n}(t)w_{n}(t) $$
we denote the total
assets  at time $t$  of funds numbered $n$ and $n-1$.

\begin{Example}
Suppose at time $t=0$ there are five funds with the following number
of units and values of units:
$$
k_{1}(0)=10^{6}, k_{2}(0)=9\cdot 10^{5}, k_{3}(0)=4\cdot 10^{5},
k_{4}(0)=3\cdot 10^{5}, k_{5}(0)=2\cdot 10^{5},
$$
and
$$
w_{1}(0)=10.5, w_{2}(0)=9.4, w_{3}(0)=4.3, w_{4}(0)=5, w_{5}(0)=8.5
$$
(in PLN). At the moment $t=1$,
$$
k_{1}(1)=10^{6}, k_{2}(1)=9.2\cdot 10^{5},
k_{3}(1)=4.3\cdot 10^{5}, k_{4}(1)=3\cdot 10^{5}, k_{5}(1)=2.2\cdot 10^{5},
$$
and
$$
w_{1}(1)=10.8, w_{2}(1)=9.7, w_{3}(1)=4.4, w_{4}(1)=5.5, w_{5}(1)=8.6.
$$
At time $\tau =1$ fund No.\ 4 merge with fund No.\ 5.
The new fund has $k_{4}(1+)=6\cdot 10^{5}$
units so the value of one unit equals
$$
w_{4}(1+)={ k_{4}(1)w_{4}(1)+k_{5}(1)w_{5}(1)\over k_{4}(1+)} = 5.9.
$$
Assume that at time $t=2$:
$$
k_{1}(2)=1.2\cdot 10^{6}, k_{2}(2)=9.4\cdot 10^{5},
k_{3}(2)=4.3\cdot 10^{5}, k_{4}(2)=6.1\cdot 10^{5},
$$
$$
w_{1}(2)=10.9, w_{2}(2)=9.6, w_{3}(2)=4.4, w_{4}(2)=6.2.
$$
Then the average return on the time period $[0,2]$ equals
$$
\bar{r}_{A}(0,2)=\left(\sum^{5}_{j=1}A^{*}_{j}(0){w_{j}(1)
\over w_{j}(0)}\right)\left(\sum^{3}_{j=1}A^{*}_{j}(1){w_{j}(2)
\over w_{j}(1)}+A^*_{4}(1){w_{4}(2)\over w_{4}(1+)}\right) - 1,
$$
(see \eqref{(2.18)}, so $\bar{r}_{A}(0,2)=0.040384=4\%$.

By \eqref{(2.19)} the rate of return of the new fund at $\tau =1$ equals 5.312\%.
Observe that the arithmetic mean of the rate of return of fund No.\ 4
and fund No.\ 5 at time $\tau =1$ is equal to $(10\%+1.117\%)/2=5.58\%$ and
is greater than the rate of return of the new fund at the same time.
\end{Example}

\end{document}